\documentclass{aastex6}
\shortauthors{Kusunose \& Takahara}

\begin{document}

\title{A Photo-Hadronic Model of the Large Scale Jet of PKS 0637-752}

\author{Masaaki Kusunose\altaffilmark{1}}
\affil{Department of Physics, School of Science and Technology,
  Kwansei Gakuin University, Gakuen 2-1, Sanda, Hyogo, 669-1337, Japan}

\and 

\author{Fumio Takahara\altaffilmark{2}}
\affil{Department of Earth and Space Science, Graduate School of Science,
  Osaka University,
  Machikaneyama 1-1, Toyonaka, Osaka 560-0043, Japan}

\altaffiltext{1}{kusunose@kwansei.ac.jp}
\altaffiltext{2}{takahara.fumio@wine.plala.or.jp}

\begin{abstract}
  Strong X-ray emission from large scale jets of radio loud quasars still remains 
  an open problem. 
  Models based on inverse Compton scattering off CMB photons by relativistically 
  beamed jets have recently been ruled out,  since \textit{Fermi} LAT observations for 
  3C 273 and PKS 0637-752 give the upper limit far below the model prediction.
  Synchrotron emission from a separate electron population with multi-hundred TeV 
  energies remains a possibility although its origin is not well known. 
  We examine a photo-hadronic origin of such high energy electrons/positrons, 
  assuming that protons  are accelerated up to $10^{19}$ eV 
  and produce electrons/positrons through  
  Bethe-Heitler process and photo-pion production. 
  These secondary electrons/positrons are injected at sufficiently 
  high energies and produce 
  X-rays and $\gamma$-rays by synchrotron radiation
  without conflicting with the \textit{Fermi} LAT upper limits. 
  We find that the resultant spectrum well reproduces the X-ray observations from 
  PKS 0637-752,
  if the proton power is \replaced{as large as}{at least}  
  $10^{49} \, \text{erg} \,\,\text{s}^{-1}$, 
  \added{which is highly super-Eddington.}
  It is noted that the X-ray emission originates primarily from leptons through 
  Bethe-Heitler process, while leptons from photo-pion origin lose energy directly through 
  synchrotron emission of multi-TeV photons rather than cascading. 
  To avoid the overproduction of the optical flux, 
  optical emission is primarily due to synchrotron emission of secondary leptons 
  rather than primary electrons, or a mild degree of beaming of the jet is needed,
  \added{if it is owing to the primary electrons.}
  Proton synchrotron luminosity is a few orders of magnitude smaller.
\end{abstract}


\section{Introduction} \label{sec:intro}

\textit{Chandra} X-ray observatory has found strong X-ray emission from large scale jets of 
many radio loud quasars. Some of them, most notorious, PKS 0637-752 
\citep{schwartz2000,chartes2000} and 3C 273 \citep{sambruna2001,marshall2001}, 
are hard to explain 
by synchrotron radiation from a high energy extension of radio-optical 
emitting electrons, 
since the observed X-ray flux is far above the extension from radio-optical 
spectra and since the X-ray spectrum is harder than the optical one.
The most conventional synchrotron self-Compton (SSC) model requires 
a very small magnetic field strength,
which means a large departure from equi-partition between 
magnetic field and relativistic electrons with an enormous jet power.
The latter has been regarded unlikely. 
It is then considered that inverse Compton (IC) scattering of cosmic microwave 
background (CMB) photons may explain the X-ray emission provided that the jet is 
relativistic and makes a small angle with the line of 
sight \citep{tavecchio2000,cgc2001}.
\deleted{In this model, however, the intrinsic jet length becomes as large as Mpcs, 
rarely seen in the parent population of radio galaxies. }
Both SSC and IC/CMB models predict that high energy extension of the 
X-ray spectrum reaches GeV energy range which can be detected with \textit{Fermi} LAT 
\citep{georganopoulos2006}. 
Recently,  \textit{Fermi} LAT observations have been reported to put upper limits of 
the $\gamma$-ray flux for the jets of 3C 273 \citep{mg2014}
and PKS 0637-752 \citep{meyer2015}. 
They are an order of magnitude lower than the SSC and IC/CMB predictions  
so that these models are incompatible with observations. 

In this situation, several alternative models should be considered. 
Within the leptonic scenario, a separate population of electrons from 
those emitting radio-optical photons, extending up to around 100 TeV is considered. 
Those electrons emit X-ray photons by synchrotron radiation. 
The existence of such an electron 
population is rather ad-hoc although not impossible. 
Its origin is usually considered a separate particle acceleration process 
from that for the radio-optical emitting population.
However, it is hard to imagine such an efficient acceleration mechanism, 
since the high energy end of radio-optical 
emitting electrons is determined by the balance of acceleration and radiative cooling. 
Actually, for PKS 0637-752, radio, optical, and X-ray emissions are spatially coincident and 
concentrated in a few bright knots, the size of which is an order of kpc, 
and the broadband spectra imply a break around $10^{12}$-$10^{13}$ Hz which 
is naturally due to radiative cooling across the source.

Alternatively, relativistic hadrons may be responsible for the X-ray emission.
\cite{bg2016} and earlier \cite{aharonian2002} considered proton synchrotron emission 
assuming magnetic field stronger than 10 mG and a large energy density of 
accelerated protons up to more than 10 PeV. 
Although they argued that the proton and Poynting powers are modest, it is due to 
their adopted time scale. They assumed that the confinement time is as long as 
$10^7$-$10^8$ years,
which is a few orders of magnitude longer than the light 
crossing time of $10^3$-$10^4$ years. 
Since the jet does not involve a large inertia nor a large confining pressure, 
we regard that the relevant time scale should be similar to the latter. 
Then, the required proton and Poynting powers become 
an order of $L_p \approx 10^{50}$ erg s$^{-1}$ 
and $L_\mathrm{Poy} \approx 10^{49}$ erg s$^{-1}$, respectively.
At the same time, the energy density of relativistic electrons becomes
about six orders of magnitude smaller than that of the Poynting power. 

Relativistic protons also contribute to X-ray and $\gamma$-ray emission 
through the production of high energy electrons and positrons by
Bethe-Heitler and photo-pion processes. 
The latter processes were considered before the \textit{Chandra} era by \cite{mkb91}
for a possible mechanism of the X-ray emission from hot spots of radio galaxies.
\cite{aharonian2002} also examined these processes for the knot A1 of 3C 273 and 
concluded that they are too inefficient.
But, the efficiency of these processes depends on the size of the knot as well as 
the infrared photon spectrum.
\added{In fact, the energy density of mm and sub-mm radiation 
in \cite{aharonian2002} is smaller than our model presented below.}
The proton models have been discussed for more compact emission regions of blazars and 
Bethe-Heitler process can contribute to the same order, depending on the 
soft photon spectral shape \citep{pm2015}.
Thus, in this paper, we consider both Bethe-Heitler and photo-pion processes  
and examine if these processes can explain the X-ray emission from PKS 0762-752.

In Section \ref{sec:estimate} we make a rough estimate of physical quantities 
of the X-ray emitting large scale jet of PKS 0637-752, whose redshift is 0.651. 
In Section \ref{sec:model} we formulate the problem, 
and in Section \ref{sec:results} numerical results are presented. 
In Section \ref{sec:conclusion} we draw conclusions.

\section{Rough Estimate}  \label{sec:estimate}

We first make a rough estimate of the physical quantities in order to 
capture the essence of the problem. For simplicity, 
we assume a single  uniform sphere of radius $R=R_\mathrm{kpc}$ kpc for the emission region
and ignore effects of relativistic beaming and redshift for the time being. 
Although the emission region is divided into a few knots in reality, 
we here treat a combined emission region.

Observed spectra at radio through optical frequencies suggest that 
the synchrotron emission from primary electrons has a peak at infrared band 
around $10^{12}$ Hz with a luminosity $L_\mathrm{syn}$ about 
$3\times 10^{44}$ erg s$^{-1}$. 
Thus,  energy density of synchrotron photons $u_\mathrm{syn}$ is about 
\begin{equation}
  u_\mathrm{syn}=\frac{3L_\mathrm{syn}}{4\pi R^2c}
  \approx 3 \times 10^{-10}R_\mathrm{kpc}^{-2} \,\,  \text{erg} \,\, \text{cm}^{-3} .
\end{equation}
The number density of photons at radio frequencies may be approximated by 
\begin{equation}
  \nu n_\nu\approx 10^6 \left(\frac{\nu}{10^{10} \, \text{Hz}}\right)^{-0.75}
  R_\mathrm{kpc}^{-2} \,\, \text{cm}^{-3} 
\end{equation}
and that at optical frequencies 
\begin{equation}
  \nu n_\nu\approx 10^{2}\left(\frac{\nu}{10^{14} \, \text{Hz}}\right)^{-1.25}
  R_\mathrm{kpc}^{-2} \,\, \text{cm}^{-3} .
\end{equation}
For the magnetic field strength of $B=B_\mathrm{mG}$ mG, 
the energy density of magnetic field is 
\begin{equation}
  u_\mathrm{mag}=4 \times 10^{-8} B_\mathrm{mG}^2 \,\, \text{erg} \,\, \text{cm}^{-3} ,
\end{equation}
and the Poynting power is estimated as  
\begin{equation}
  L_\mathrm{Poy}=\pi R^2 u_\mathrm{mag}c 
  \approx 4\times 10^{46}B_\mathrm{mG}^{2}R_\mathrm{kpc}^{2} 
  \,\, \text{erg} \,\, \text{s}^{-1} .
\end{equation}

The Lorentz factor of electrons ranges from 
\replaced{$\gamma_{e , \mathrm{min}}\approx 3 \times 10^3B_\mathrm{mG}^{-0.5}$}
{$\gamma_{e , \mathrm{min}}\approx 2 \times 10^3B_\mathrm{mG}^{-0.5}
 (\nu_\mathrm{min}/10^{10} \, \mathrm{Hz})^{0.5}$}
to $\gamma_{e, \mathrm{max}}\approx 10^6B_\mathrm{mG}^{-0.5}$
with a broken power law spectrum. 
The power law index of electrons is tentatively taken as 2.5 at low energies 
and 3.5 at high energies in accordance with the above photon spectra. 
\deleted{The energy density of electrons is governed by the low energy end}
\added{The peak luminosity of $3 \times 10^{44}$ erg s$^{-1}$  at $10^{12}$ Hz
is emitted by primary electrons with 
$\gamma_\mathrm{br} \sim 2 \times 10^{4} B_\mathrm{mG}^{-0.5}$ below which 
the number spectrum is given by $n_e(\gamma_e) = K_e \gamma_e^{-2.5}$.
On the other hand, $L_\mathrm{syn}$ is given by 
$\sim (4 \pi R^3/3) c \sigma_\mathrm{T} u_\mathrm{mag} \gamma_\mathrm{br}^2 
n_e(\gamma_\mathrm{br}) \gamma_\mathrm{br}$,
where $\sigma_\mathrm{T}$ is the Thomson cross section.
From these relations we write $K_e$ in terms of $\gamma_\mathrm{br}$, $R$, and $B$.
The electron energy density is now given by
$u_e \sim 2 m_e c^2  K_e \gamma_\mathrm{min}^{-0.5}$ 
}
and estimated as  
\begin{equation}
  u_e \approx  8  \times 10^{-10} B_\mathrm{mG}^{-1.5} R_\mathrm{kpc}^{-3}
  \left(\frac{\nu_\mathrm{min}}{10^{10} \mathrm{Hz} }\right )^{-0.25} 
  \,\, \text{erg} \,\, \text{cm}^{-3} .  
  \label{eq:ue}
\end{equation}
It may seem that \replaced{inverse Compton}{SSC} luminosity can be large,
if the magnetic field strength is smaller than 0.1 mG for a typical source size of 1 kpc. 
However, in this case, most of the inverse Compton 
emission is produced in the  MeV-GeV range. 
To reproduce the observed X-ray flux, magnetic field needs to be as small as 0.01 mG.
In this case, the kinetic power of electrons given by
\begin{equation}
  L_e= \frac{4\pi R^3u_e}{3}\frac{c}{3R}
  \approx  3 \times  10^{44} B_\mathrm{mG}^{-1.5} R_\mathrm{kpc}^{-1}
  \left(\frac{\nu_\mathrm{min}}{10^{10} \text{Hz} }\right)^{-0.25}
  \,\, \text{erg} \,\, \text{s}^{-1} 
\end{equation}
would become very large, 
\added{i.e., $L_e \sim 3 \times 10^{47}$ erg s$^{-1}$ for $B = 0.01$ mG
and $\nu_\mathrm{min} = 10$ GHz.}
Here, we take the escape time of electrons as $3R/c$. 
Historically, for this reason, the beamed IC/CMB model was proposed, 
but as noted in Section \ref{sec:intro}, this model is now regarded unlikely. 
The minimum power for explaining the radio-optical flux 
is realized at $B_\mathrm{mG}\approx 0.2$ for $R_\mathrm{kpc}=1$
with $L_\mathrm{Poy} \approx L_e\approx 2 \times 10^{45} \, \text{erg} \,\, \text{s}^{-1}$.

\added{Theoretically,} 
the break energy of the electron energy distribution, \added{$\gamma_b$,} 
is determined by the 
balance between synchrotron cooling and escape; if we equate the cooling time 
with escape time, we obtain
\begin{equation}
  \gamma_b \approx 3 \times 10^3B_\mathrm{mG}^{-2}R_\mathrm{kpc}^{-1} .
\end{equation}
If the break corresponds to the break of radio-optical spectrum at $10^{12}$ Hz, 
we obtain the field strength of around 0.2 mG for $R=1 \, \text{kpc}$.
These considerations lead to $B_\mathrm{mG}\approx 0.1-0.3$ as an appropriate choice.
 
The maximum possible Lorentz factor of electrons is estimated by equating the 
cooling time with the  
gyrotime and given by 
\begin{equation}
  \gamma_{e, \mathrm{lim}} \approx 10^9 B_\mathrm{mG}^{-1/2}  .
\end{equation}
Thus, in principle it is possible to obtain $\gamma_e$ 
as large as $10^8$ with which electrons emit synchrotron X-rays. 
However, since the observed X-ray spectrum is much flatter than the optical spectrum, 
such a high energy population should be separate from radio-optical emitting one and
the acceleration mechanism should be very efficient and distinct. 
Alternatively, such electrons may be supplied from photo-hadronic processes.
It should be noted that AGN jets are composed of protons and electrons/positrons and 
the inertia is likely to be dominated by protons \added{\citep{uchiyama2005}}, 
while the existence of electron-positron pairs is also suggested
by various analysis of observations. 
Proton acceleration \added{may} also naturally \deleted{takes} \added{take} place 
and in principle the maximum energy of 
protons can be as large as  $10^{20}$ eV for 1 mG field and 1 kpc size. 

Two photo-hadronic processes can provide secondary high energy electrons/positrons, i.e., 
photo-pion production process and Bethe-Heitler process. 
The former is through strong interaction with the cross section of about 
$3\times 10^{-28} \,\, \mathrm{cm}^2$ 
and the threshold energy of
\begin{equation}
  \gamma_{p, \mathrm{th}} \approx m_\pi c^2 \epsilon_\mathrm{soft}^{-1}
  = 3\times 10^{12} \left(\frac{\nu_\mathrm{soft}}{10^{10} \mathrm{Hz}}\right)^{-1},
\end{equation}
where $\epsilon_\mathrm{soft}=h \nu_\mathrm{soft}$ is the energy of a target photon. 
Thus, for $\gamma_p=10^{10}$, the energy of target photons should be larger than 
$3\times 10^{12}$ Hz
with the number density about  $10^5 \,\, \text{cm}^{-3}$ for $R_\mathrm{kpc}=1$. 
Thus, a proton interaction probability is around 0.1 for rectilinear propagation. 
Charged pions decay to produce electrons and positrons, while neutral pions decay into 
two $\gamma$-rays, which interact with soft photons to produce electron-positron pairs. 
\added{About 5 \% of the inelastic energy goes into electron/positrons.
This is because pion mass is about 15 \% of proton mass, so that about 15 \% of
proton energy goes to pions near the threshold.
This energy is further distributed to 4 leptons almost equally,
\cite[e.g.,][]{ka2008,dm2009}.}
\deleted{Considering that about 5 \% of the inelastic energy goes into electron/positrons,  }
\added{Considering this,}
we estimate 0.5 \% of the proton power can be used to produce electrons 
and positrons with the Lorentz factor of around $2.5 \times 10^{11}$, 
which subsequently radiate synchrotron radiation below the TeV energy 
region for $B=1 \,\text{mG}$. 
While photons with energy higher than TeV is optically thick to 
photon-photon pair production, most synchrotron photons are emitted below TeV and 
pair cascade process does not much develop. 
Since these electrons/positrons rapidly cool to make 
the energy distribution of a power law with an index of $-2$, 
the resultant photon energy spectrum is a power law with an index of $-0.5$
and the X-ray  luminosity is four orders of magnitudes smaller than the TeV luminosity. 
Thus, if we would explain the X-ray observations with this mechanism, 
the predicted GeV luminosity becomes around 
$3 \times 10^{47} \, \text{erg} \,\, \text{s}^{-1}$, 
which exceeds the \textit{Fermi} LAT upper limit. 
Required proton power is uncomfortably large 
amounting to $3 \times 10^{51} \, \, \text{erg} \,\, \text{s}^{-1}$.
Higher energy protons exacerbate the problem and lower energy protons 
have a negligibly small interaction probability.

Bethe-Heitler process has a larger cross section of 
$3\times 10^{-27} \, \text{cm}^2$ and the lower threshold energy of 
\begin{equation}
  \gamma_{p, \mathrm{th}}= m_e c^2\epsilon_\mathrm{soft}^{-1}
  = 3\times 10^{10} \left(\frac{\nu_\mathrm{soft}}{10^{10} \, \mathrm{Hz}}\right)^{-1} ,
\end{equation}
but with a lower efficiency of energy transfer to electrons and positrons.
For $\gamma_p =10^{10}$, target photon energy is larger than $3\times 10^{10}$ Hz 
with the number density $5\times 10^5 \,\, \text{cm}^{-3}$ for $R_\mathrm{kpc}=1$, 
and on average a proton produces 5 electron-positron pairs before it escapes from the region.
Taking the efficiency of 0.001, 0.5\% of the proton power can be used to 
pair production, roughly the same order as the photo-pion production. 
In this case, however, the injection Lorentz factor of electrons and positrons is around 
\deleted{$10^{10}$} \added{$\gamma_e \sim \gamma_p = 10^{10}$} 
and the synchrotron frequency is peaked at 100MeV for $B=1$ mG. 
Since the resultant synchrotron spectrum becomes a power law with an index of $-0.5$,
the X-ray luminosity  is about two orders of magnitudes lower than the 100 MeV 
luminosity. 
The required proton power is around $10^{49} \,\, \text{erg} \, \, \text{s}^{-1}$, 
which is very large but not inconceivable considering the Poynting power is 
an order of $4 \times 10^{46} \,\, \text{erg} \,\, \text{s}^{-1}$. 
When the magnetic field strength is 0.1 mG, 
the required proton power is $3\times 10^{48} \,\, \text{erg} \,\, \text{s}^{-1}$,
\added{because the beak frequency of synchrotron radiation by secondary leptons
decreases as $B$ decreases.}
The power of primary electrons becomes 
 $10^{46} \,\, \text{erg} \,\, \text{s}^{-1}$ while the Poynting power 
is $4 \times 10^{44} \,\, \text{erg} \,\, \text{s}^{-1}$. 
Thus we regard about $B = 0.1 \,\text{mG}$ is a best guess. 
\added{The proton power $\sim 10^{49}$ erg s$^{-1}$ estimated above
  is $\sim 120 L_\mathrm{Edd}$,
  where $L_\mathrm{Edd}$ is the Eddington luminosity with the black hole mass
  $M_\mathrm{BH} = 6.5 \times 10^8 M_\sun$ \citep{ljg2006}.
  It is to be noted that $M_\mathrm{BH} = 7.8 \times 10^9 M_\sun$ has been 
  reported by \cite{gcj2001}.  For this value of $M_\mathrm{BH}$, 
  the proton power is $\sim 10 L_\mathrm{Edd}$.}

Somewhat lower energy protons also contribute to Bethe-Heitler process; although the 
interaction probability becomes lower, the injection Lorentz factor also becomes smaller, 
which makes the synchrotron peak lower.
For example, for $\gamma_p=10^9$, the target photon energy is above $3\times 10^{11}$ Hz
with the number density of $10^5 \,\, \text{cm}^{-3}$, 
and the interaction probability of a proton becomes 1. 
Thus 0.1\% of the proton power is available. 
The injection Lorentz factor of electrons/positrons is also $10^9$, which emit 
synchrotron radiation at 1 MeV for $B=1 \, \text{mG}$ or 0.1 MeV for $B=0.1 \, \text{mG}$.
So the proton power of $3 \times 10^{48} \,\, \text{erg} \,\, \text{s}^{-1}$ 
gives rise to the observed X-ray luminosity. 

It is to be noted that for a larger $R_\mathrm{kpc}$ the interaction probability 
of protons becomes small and required proton power accordingly increases; 
a larger $R_\mathrm{kpc}$ is unfavorable for the photo-hadronic model. 
For these parameters, proton synchrotron luminosity is a few orders of magnitude 
smaller than the synchrotron luminosity of secondary pairs.  
Its peak is at $10^{19}$ Hz for $B=0.1\, \text{mG}$.

Since the resultant emission spectra depend on details of the secondary 
electron/positron spectra and resultant radiative cooling, 
we numerically investigate such details in the next section. 
Based on the present investigation, in this paper, 
we concentrate on Bethe-Heitler process.

\section{The Model} \label{sec:model}

We adopt a single zone model in which the size of the emission region of $R$
around 1 kpc, magnetic field of $B$ around 0.1 mG, and the proton spectrum is 
\begin{equation}
  n_p \added{(\gamma_p)} = K_p \gamma^{-p} ,
  \label{eq:proton-spec}
\end{equation}
for $\gamma_{p, \mathrm{min}}\le \gamma_p \le \gamma_{p, \mathrm{max}}$. 
Canonically we take $p=2$, $\gamma_{p, \mathrm{min}}=1$,
and $\gamma_{p, \mathrm{max}}=10^{10}$. 
Since photo-hadronic processes work only for large values of $\gamma_p$, 
the value of $\gamma_{p, \mathrm{min}}$ does not affect the resultant 
spectrum but only affects the energy density of protons \added{logarithmically}.
If  proton energy distribution is concentrated in the range near $\gamma_{p, \mathrm{max}}$,
the power requirement below  will be relieved by an order of magnitude.  
Note that the estimates of the proton power and proton energy density given in the 
previous section are those for such a high energy population 
and an order of magnitude lower than those for $\gamma_{p, \mathrm{min}}=1$.
In contrast $\gamma_{p, \mathrm{max}}$ critically affects the results.
The energy density of protons is 
\begin{equation}
  u_p =10^{-3} K_p \ln(\gamma_{p, \mathrm{max}}/\gamma_{p, \mathrm{min}}) \,\,
  \text{erg} \,\, \text{cm}^{-3} .
\end{equation}
The proton power is 
\begin{equation}
  L_p = \frac{4\pi R^3 u_p}{3}\frac{c}{3R}
  \approx 4 \times 10^{50} K_p
  \ln(\gamma_{p, \mathrm{max}}/\gamma_{p, \mathrm{min}}) R_\mathrm{kpc}^2 
  \,\, \text{erg} \,\, \text{s}^{-1},
\end{equation}
where we take the escape time of $3R/c$.

The target photon spectra of photo-pion and Bethe-Heitler processes 
are based on the observed radio to optical photons,
taking into account of the cosmological redshift of 0.651.
Primary electrons responsible for the  radio-optical emission are 
determined so as to reproduce the radio-optical spectra. 
We also calculate the inverse Compton scattering of primary electrons 
off radio-optical photons, 
whose flux is generally orders of magnitude short of X-ray observations.
In numerical calculations we solve the kinetic equation of primary electrons
with the injection spectrum given by 
$q_\mathrm{inj}(\gamma_e) = K_e \gamma_e^{-\alpha_e} \exp(-\gamma_e/\gamma_{e, 0})$,
where $\gamma_e$ is the Lorentz factor of electrons and $K_e$, $\alpha_e$, 
and $\gamma_{e, 0}$ are parameters to fit the observed radio-optical spectrum.
Since we consider only mildly relativistic beaming if any, we ignore IC/CMB.  
Although the number density of CMB photons is larger than  
that of radio-optical synchrotron photons when $R_\mathrm{kpc}$ is larger than 10, 
the efficiency of those processes becomes small for large $R$, 
so that our treatment is justified.

The emission by high energy leptons produced by the hadronic processes is 
calculated for  the lepton spectrum
obtained by solving the kinetic equation given by
\begin{equation}
  \frac{d n_e (\gamma_e)}{d t}=q_\mathrm{BH}(\gamma_e) + q_\mathrm{pp}(\gamma_e)
  -\frac{c n_e (\gamma_e)}{3R}
  -\frac{d~}{d\gamma_e} [\dot{\gamma}_e n_e (\gamma_e)] ,
  \label{eq:e-kinetic-hadronic}
\end{equation}
where $n_e(\gamma_e)$ is the lepton density per unit interval of $\gamma_e$.
The lepton injection rate is denoted by $q_\mathrm{BH}(\gamma_e)$ for Bethe-Heitler process, 
which is calculated according to the formulation  given by \cite{ka2008},
\added{using a given proton spectrum with equation (\ref{eq:proton-spec}).}
Lepton production via photo-pion processes is denoted by $q_\mathrm{pp}(\gamma_e)$.
However, we do not include this term in numerical calculations 
because the leptons produced by photo-pion processes do not contribute much 
to the X-ray emission.
Here $\dot{\gamma}_e$ denotes radiative cooling through synchrotron radiation 
and inverse Compton scattering.
We also set the lepton escape time to be $3 R/c$.
Using the steady solution of equation (\ref{eq:e-kinetic-hadronic}), 
we calculate the emission spectra through synchrotron radiation 
and inverse Compton scattering.

\section{Results}  \label{sec:results}

\begin{figure}[ht!]   
  \centering\includegraphics[scale=0.5,clip]{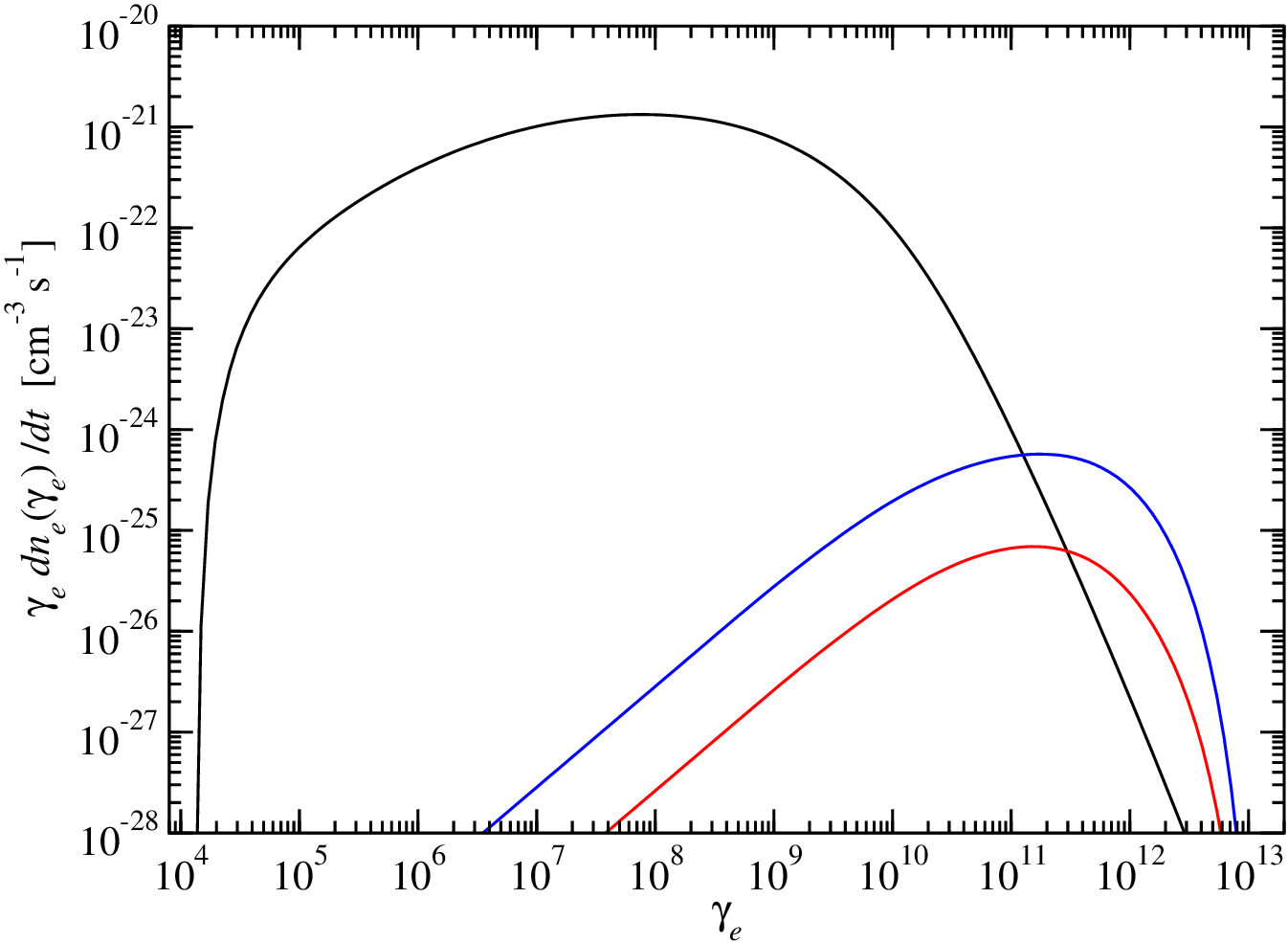}
  \caption{The production spectrum of electrons and positrons 
    for \added{$p = 2$ and} $K_p = 1$ \added{cm$^{-3}$.}
    The synchrotron radiation spectrum by primary electrons 
    for $R = 1$ kpc and $B = 0.1$ mG (Fig. \ref{fig:1kpc-no-beaming-exp})
    is used as a target photon spectrum in lepton production.
    The black line shows the pair production rate by Bethe-Heitler process,
    the red and blue lines show the electron and positron production rates, 
    respectively, by photo-pion process.
    \label{fig:pair-production-spec}}
\end{figure}

\begin{figure}[ht!]   
  \centering\includegraphics[scale=0.5,clip]{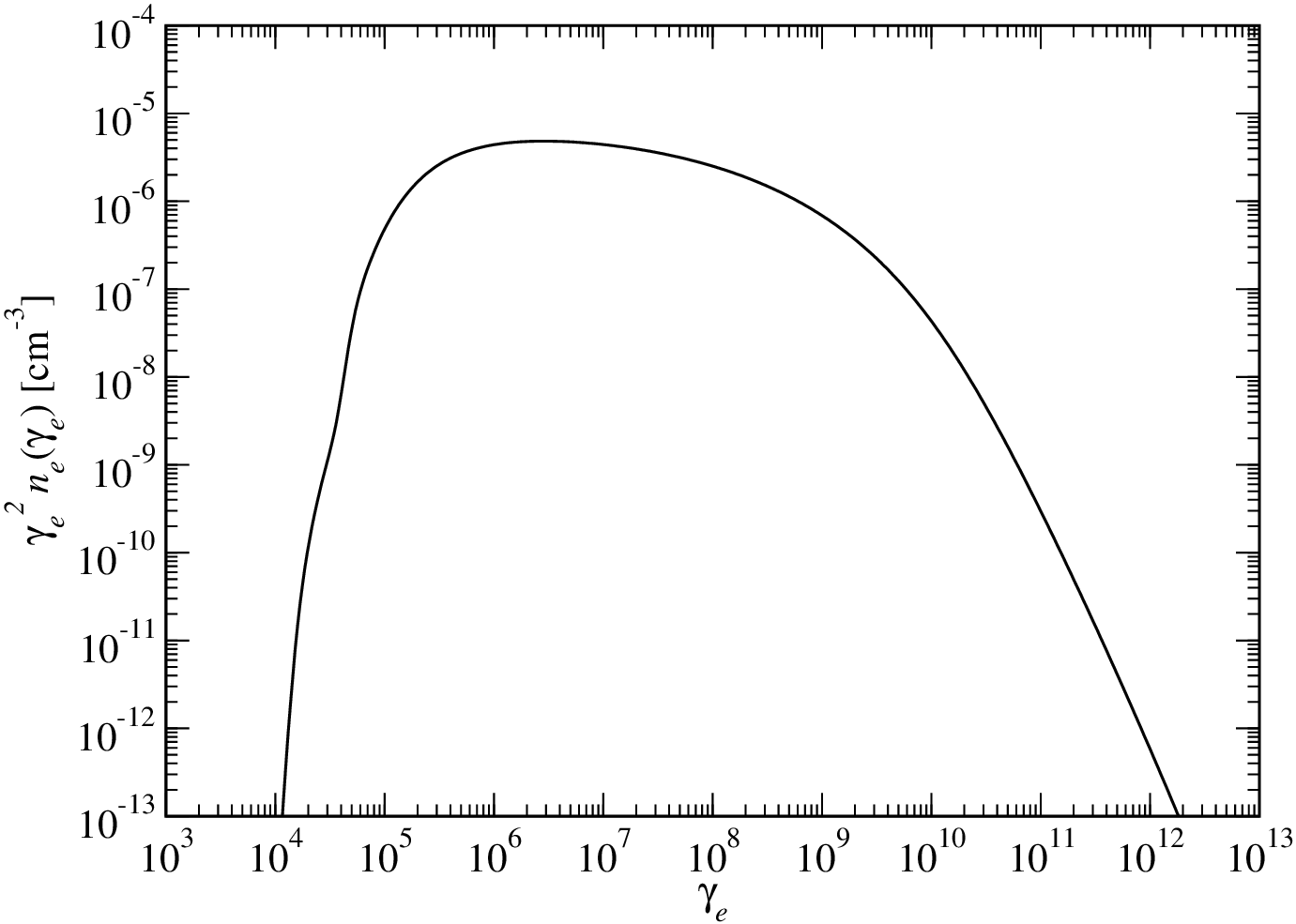}
  \caption{The lepton energy spectrum for $R = 1$ kpc, $B = 0.1$ mG,
    \added{and $p=2$}.
    The spectrum is calculated by equation (\ref{eq:e-kinetic-hadronic})
    for a steady state.
    The injection of leptons is by Bethe-Heitler process.
    \label{fig:lepton-energy-spec}}
\end{figure}

\begin{figure}[t!] 
  \centering\includegraphics[scale=0.5,clip]{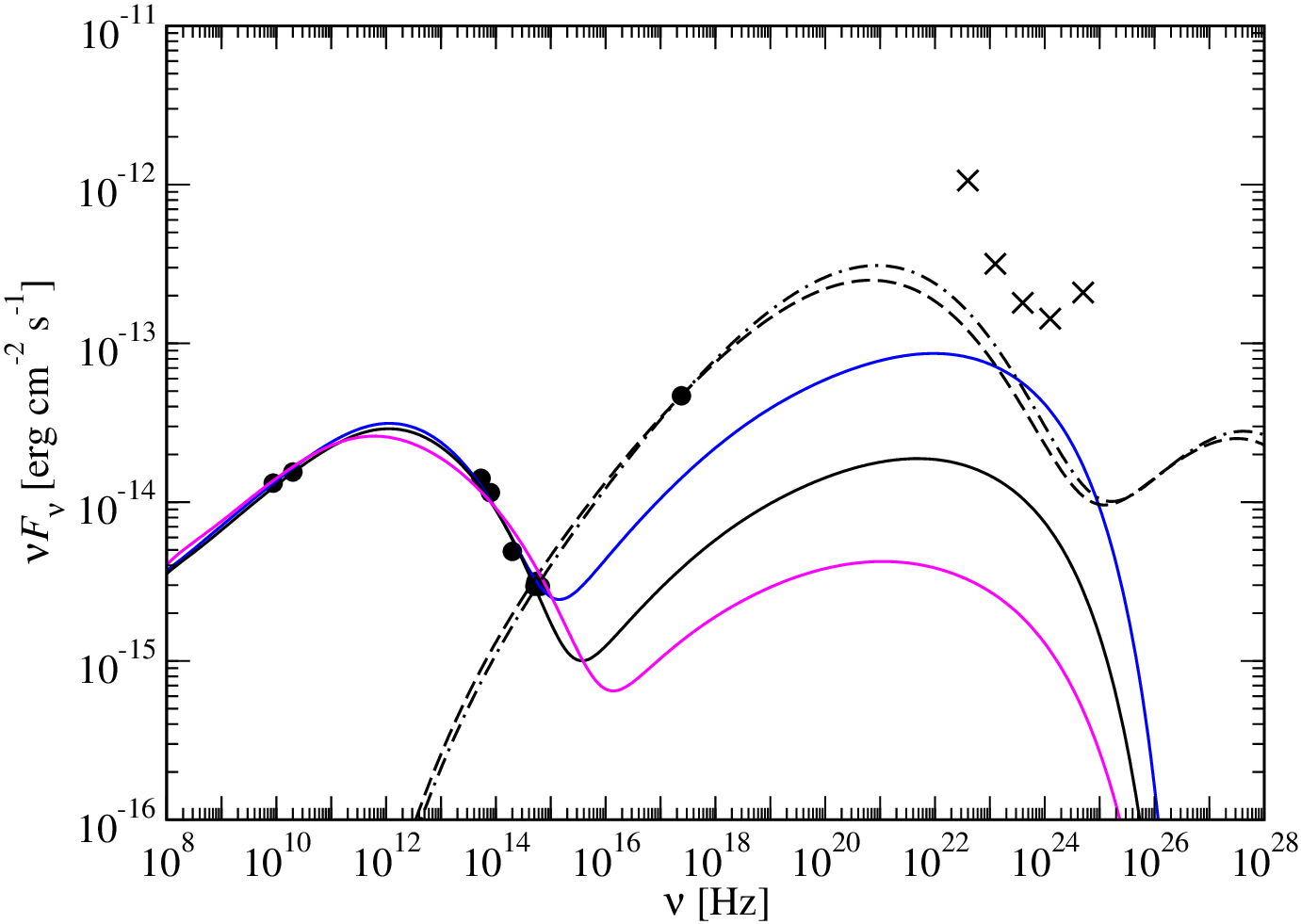}
  \caption{The emission spectrum of PKS 0637-0752.
    The black filled circles and crosses are data taken from \cite{meyer2015} 
    and the crosses show the \textit{Fermi} upper limit.
    Models are calculated for $R =1$ kpc.
    \added{Solid lines are synchrotron radiation and SSC by primary electrons.
      The bumps above $\sim 10^{15}$ - $10^{16}$ Hz of the solid lines are SSC components.}
    The blue solid line is for $B=0.05$ mG, the black solid line is 
    for $B=0.1$ mG, and the magenta solid line for $B=0.2$ mG.
    \added{Dashed and dot-dashed lines are emission by secondary pairs.}
    The \deleted{black} dashed line is for $p=2$ and $K_p = 8.8\times 10^{-3}$ 
    \added{cm$^{-3}$}
    and the \deleted{black} dot-dashed line is for $p=1.8$ and $K_p = 1.3 \times 10^{-4}$
    \added{cm$^{-3}$}.
    These spectra \added{(dashed and dot-dashed)} are calculated with $B=0.1 \,\text{mG}$.
    The bumps above $\sim 10^{26} \, \text{Hz}$ are produced by inverse Compton scattering 
    of radio-optical photons off the secondary pairs produced by the Bethe-Heitler process.
    \label{fig:1kpc-no-beaming-exp}}
\end{figure}

In Figure \ref{fig:pair-production-spec} we show the production spectrum of electrons 
and positrons through Bethe-Heitler and photo-pion processes
for a photon field relevant to PKS 0637-752 jet for the fiducial case $R_\mathrm{kpc} = 1$
and $B_\mathrm{mG}  = 0.1$ with \added{$p =2$} and $K_p=1$ \added{cm$^{-3}$}.
\added{The pair production rate is calculated based on \cite{ka2008}.
The target photons are synchrotron radiation by primary electrons,
the spectra of which are shown in Figure \ref{fig:1kpc-no-beaming-exp} 
for various parameters.}
\added{Figure  \ref{fig:pair-production-spec} and Table \ref{table:prod-rate-2-18} below
do not include pair production via the decay of neutral pions.
The gamma-rays produced by the decay of neutral pions have energy comparable to
the energy of electrons/positrons produced by charged pions.
In collisions with soft photons the gamma-rays produce electron-positron pairs, 
and these pairs contribute mainly to TeV emission.}
As we described in Section \ref{sec:estimate}, the production spectrum
by Bethe-Heitler process has a rather broad number spectrum 
centered on $10^6<\gamma_e<10^{10}$,
while those through photo-pion production have a \deleted{peaked spectrum}
\added{peak} at $\gamma_e \approx 10^{11}$-$10^{12}$.
The energy injection rate for both processes concentrates on the high energy ends,
with the Bethe-Heitler process being an order of magnitude larger than that for the
photo-pion production.
\added{For $\gamma_e \gtrsim 10^{11.5}$, 
photo-pion processes dominate the Bethe-Heitler process.
Leptons with $\gamma_e \gtrsim 10^{11.5}$ emit synchrotron radiation at 
$\nu \gtrsim 10^{25}$ Hz and do not contribute to X-ray emission that is our
interest in this paper.}

The resultant steady state electron/positron energy spectrum is shown 
in Figure  \ref{fig:lepton-energy-spec}
for $R_\mathrm{kpc}=1$, $B_\mathrm{mG}=0.1$, \added{and $p=2$}.
The value of $K_p$ is taken as $8.8 \times 10^{-3}$ \added{cm$^{-3}$} to reproduce 
the observed X-ray flux.
The proton power amounts to $5 \times 10^{49} \,\, \text{erg} \,\, \text{s}^{-1}$.
Although this seems to be too large, it can be reduced by an order of magnitude 
if $\gamma_{p, \mathrm{min}}$ is large enough or if the spectral index of protons 
is less than 2, i.e., when the energy of relativistic protons is 
concentrated in the high energy end.  
We tabulate in Table \ref{table:prod-rate-2-18}
the pair production rates for $p = 2$ and 1.8.

The resultant photon spectra are shown in Figure  \ref{fig:1kpc-no-beaming-exp}.
As is seen, while an overall spectral shape is well reproduced, 
optical flux at $5 \times 10^{14}$ Hz tends to be overproduced.
At this frequency primary electrons and secondary pairs equally contribute and 
the resultant combined flux is a factor of two higher than the observed one.
This is due to radiative cooling of secondary pairs and rather a general feature.
For a range of the magnetic field strength this feature persists in the present model.
Possible way out from this problem is discussed later.
When the magnetic field becomes smaller, inverse Compton scattering 
of radio-optical synchrotron photons by primary electrons (SSC) becomes larger.
If the magnetic field is as small as 20$\mu$G, SSC can work as an X-ray emission 
mechanism. 
In this case, however, the electron power is as large 
as $3 \times 10^{47} \, \text{erg} \,\, \text{s}^{-1}$ with a large deviation 
from equi-partition.
Furthermore, excessive production of optical photons by IC is inevitable.
Thus, SSC model does not work.

The secondary pairs scatter the synchrotron photons emitted by the primary electrons.  
This IC component does not affect the emission below
$\sim 10^{24} \, \text{Hz}$ but the peak in $\nu F_\nu$ appears in the TeV band.
The IC component is not much affected by the magnetic field strength
because the spectra of the  (radio-optical) soft photons and the secondary pairs are fixed.
It is \deleted{, however,} to be noted that 
our model does not include the absorption of $\gamma$-rays
by extragalactic background light.
\added{The TeV bump will be hard to be observed by CTA,
because the expected flux is lower than the lower limit of CTA.}

\floattable
\begin{deluxetable}{c|rrrrrr}
  \tablecaption{Pair Production Rate for $R = 1$ kpc and  $B = 0.1$ mG
  \label{table:prod-rate-2-18}}
  \tablewidth{0pt}
  \tablehead{
    \colhead{} & \multicolumn{2}{c}{$p=2$} 
    & \colhead{} &  \multicolumn{3}{c}{$p=1.8$}
    \\
    \cline{2-3}
    \cline{5-7}
    \colhead{} &  \colhead{$\dot{n}_\pm$} &  \colhead{$q_\pm$}
    & 
    \colhead{}  &   \colhead{} &  \colhead{$\dot{n}_\pm$} &  \colhead{$q_\pm$}
    \\
    \colhead{} & \colhead{(cm$^{-3}$ s$^{-1}$)} & \colhead{(erg cm$^{-3}$ s$^{-1}$)} 
    &
    \colhead{}  &\colhead{}   & \colhead{(cm$^{-3}$ s$^{-1}$)} & \colhead{(erg cm$^{-3}$ s$^{-1}$)} 
  }
  \startdata
  B-H &  $9.4 \times 10^{-21}$ \phn & $4.0 \times 10^{-18}$  \phn 
 & &  
  &  $4.8 \times 10^{-19}$ \phn & $3.2 \times 10^{-16}$  \phn 
  \\ 
  $e^-$ &  $2.6 \times 10^{-25}$ \phn & $5.4 \times 10^{-20}$ \phn 
  & & 
  &  $1.9 \times 10^{-23}$ \phn & $3.9 \times 10^{-18}$ \phn 
  \\ 
  $e^+$ & $2.4 \times 10^{-24}$ \phn & $5.5 \times 10^{-19}$ \phn 
  & &
  &   $1.5 \times 10^{-22}$ \phn & $4.0 \times 10^{-17}$ \phn 
  \\ 
  \enddata
  \tablecomments{
    $\dot{n}_\pm$: the electron/positron production rate per unit volume,
    $q_\pm$: the energy production rate per unit volume in electron/positron production.
    B-H: Bethe-Heitler pair production, $e^-$ and $e^+$: photo-pion processes.
    $K_p = 1$ \added{cm$^{-3}$}, $\gamma_{p, \mathrm{min}}=1$,
    and $\gamma_{p, \mathrm{max}} = 10^{10}$ are assumed.
    All the values in the table are proportional to $K_p$.
    }
\end{deluxetable}

\begin{figure}[ht!]      
  \centering\includegraphics[scale=0.5,clip]{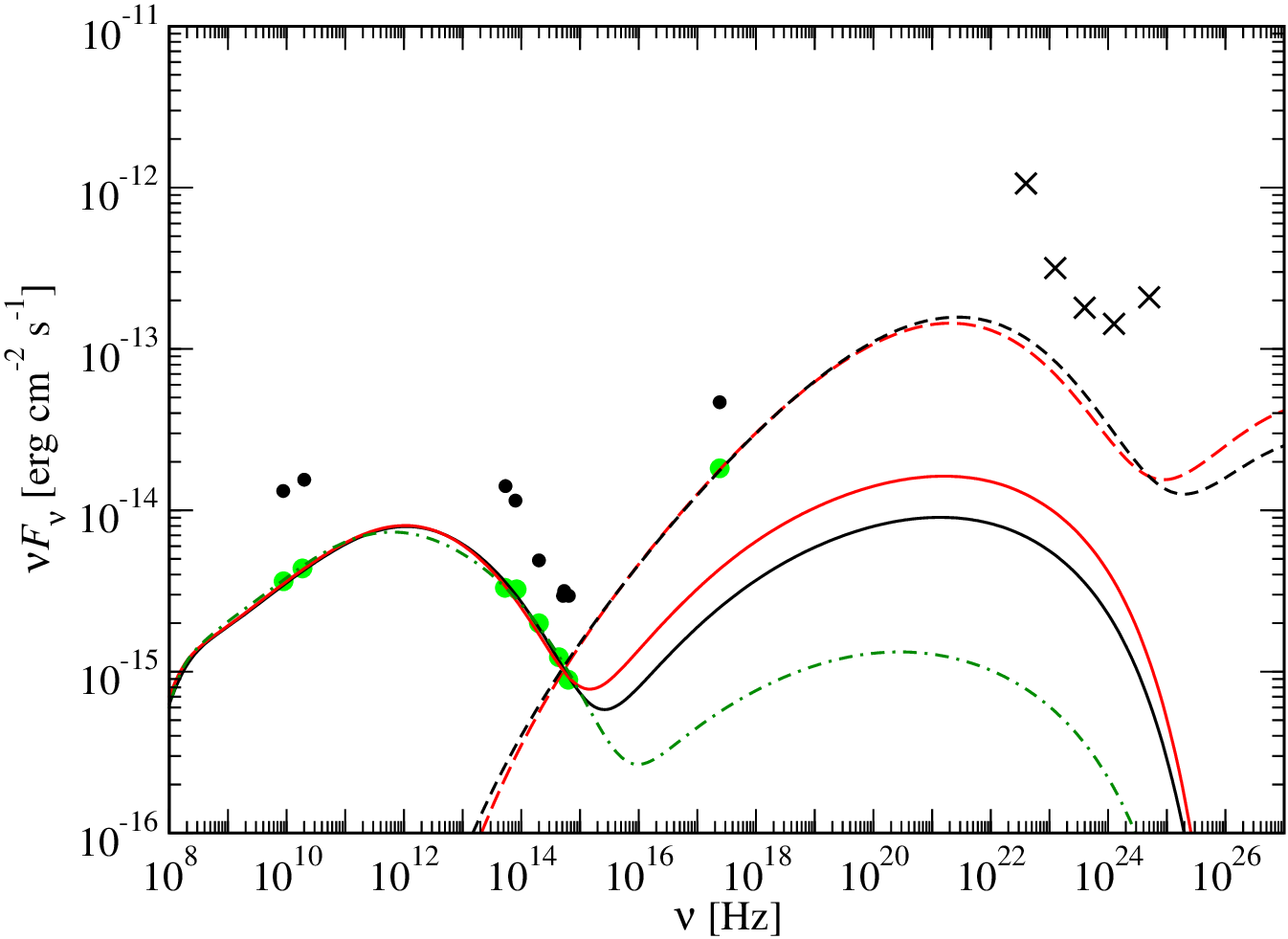}
  \caption{Emission spectrum of the knot WK8.9 (the data are shown by filled green
    circles taken from \cite{meyer2015} ).
    The size of the emission region is assumed to be $0.1$ kpc.
    The red lines \added{(solid and dashed)} are for $B = 0.3$ mG, 
    \added{and} the black lines \added{(solid and dashed)} are for $B=0.4$ mG.
    \deleted{and the green dot-dashed line is for $B=1$ mG.}
    \added{For $B = 1$ mG, only synchrotron radiation and SSC by primary electrons is shown
      by the green dot-dashed line.}
    To calculate the emission from pairs produced by Bethe-Heitler process, 
    $K_p = 0.18$ and 0.2 \added{cm$^{-3}$} are assumed for $B=0.3$ and 0.4 mG, respectively.
    \label{fig:01kpc-spec}}
\end{figure}

We also examined a smaller size of $R_\mathrm{kpc}=0.1$ and applied to 
the knot WK8.9 as shown in Figure \ref{fig:01kpc-spec}.
When the magnetic field is $B_\mathrm{mG}=0.3$, $K_p=0.18$ \added{cm$^{-3}$} can 
reproduce the radio-optical and X-ray spectra, although 
overproduction of optical photons  still persists. 
The proton power is around $10^{49} \, \, \text{erg} \,\, \text{s}^{-1}$.

For a larger value of $R$, the required proton power increases; for example 
for $R_\mathrm{kpc} = 5$, $K_p =10^{-3}$ \added{cm$^{-3}$} is needed amounting to 
$L_p =10^{50} \,\, \text{erg}\,\, \text{s}^{-1}$.
Thus, as for the emission region size, a small radius is favored in the energetics of protons. 
These numerical results are consistent with a rather simple and optimistic estimate made 
in the previous section.
The predicted photon spectra show a roll-over at 10 MeV-GeV range 
and are compatible with the reported \textit{Fermi} upper limits.
Since the spectra are rather flat, the real problem is in the low energy 
end of the synchrotron emission. 
We mostly skipped the contribution from electrons/positrons of photo-pion origin.
They contribute mostly in the TeV range
with roughly similar luminosity to X-rays so that they do not affect the X-ray spectrum 
and the \textit{Fermi} upper limit.
When the maximum energy of protons is not so large, this component 
can be totally ignored.

\begin{figure}[ht!]      
  \centering\includegraphics[scale=0.5,clip]{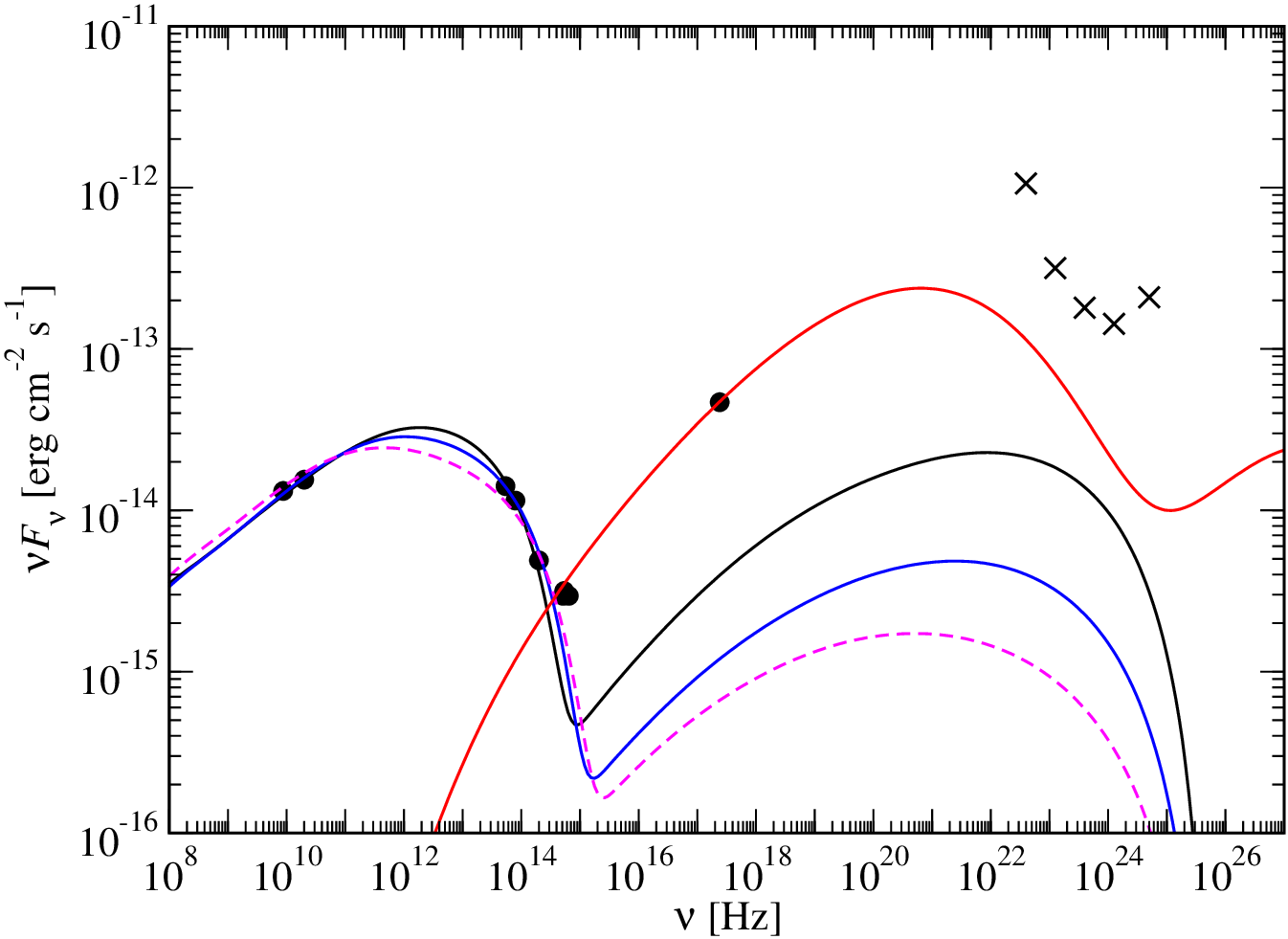}
  \caption{
    The injection of primary electrons without exponential cutoff
    is assumed for $R = 1$ kpc.
    The black line is for $B=0.1$ mG, the blue line is for $B=0.2$ mG, and the magenta line 
    is for  $B = 0.3$ mG.
    The red line is the emission from pairs produced by Bethe-Heitler pair production for
    $B= 0.1$ mG and $K_p=8 \times 10^{-3}$ \added{cm$^{-3}$}.
    \label{fig:1kpc-no-exp-cutoff}}
\end{figure}

Since the overproduction of optical flux is rather general, 
we consider the reduction of the emission from the primary electrons.
In the above models we assumed the injection spectrum with the exponential cutoff.
When we assume the power-law  injection spectrum without the exponential cutoff,
the optical emission is mainly from the secondary pairs.
Our numerical result is shown in Figure \ref{fig:1kpc-no-exp-cutoff}.
Such an abrupt super-exponential cutoff of primary electron 
energy distribution may not be unlikely,
when the acceleration is limited by cooling \citep{krm98}.

\begin{figure}  [ht!]          
  \centering\includegraphics[scale=0.5,clip]{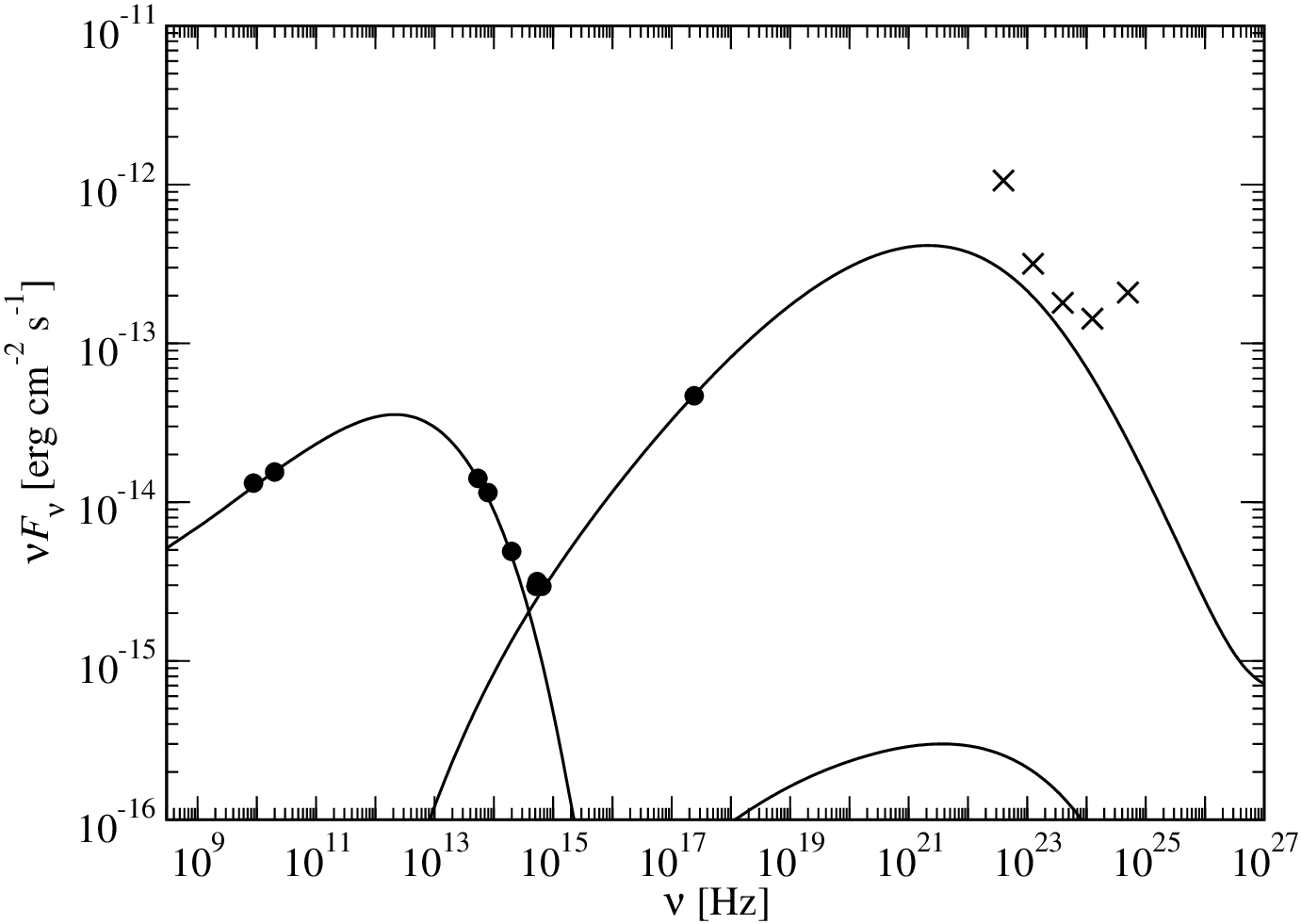}
  \caption{Beaming factor $\delta = 3$ is assumed for $R =1$ kpc and $B = 0.1$ mG.
    To calculate X-ray emission, $K_p = 1.1 \times 10^{-2}$ \added{cm$^{-3}$} is assumed.
    \label{fig:beaming-spec}}
\end{figure}

An alternative idea to reduce the overproduction of the optical flux is 
taking into account a mild relativistic beaming.
In this case, the photo-hadronic rates become less frequent
by a factor of $\delta^{3.75}$ due to the beaming effects for the same value 
of $\gamma_p$,
\added{where $\delta$ is the beaming factor.}
\added{(The scaling of $\delta^{3.75}$ is obtained as follows.
The observed frequency $\nu_\mathrm{obs}$ and the frequency in the source 
$\nu_s$ are related by $\nu_\mathrm{obs} = \delta \nu_s$. 
The photon density at $\nu_\mathrm{obs}$ is given by
$\nu_\mathrm{obs} n_{\nu_\mathrm{obs}} = \delta^3 \nu_s n_{\nu_s}
= A \nu_\mathrm{obs}^{-0.75} = A \delta^{-0.75} \nu_s^{-0.75}$, where $A$ is a constant.
Then $\nu_s n_{\nu_s} = A \delta^{-3.75} \nu_s^{-0.75}$ follows.)
}
However, the source frame X-ray luminosity also decreases by a factor of $\delta^4$, 
so that  the required proton power in the source frame does not much change,
\added{i.e., proportional to $\delta^{0.25}$.}
The required kinetic power of protons increases by a factor of $\delta^2$,
if we set the bulk Lorentz factor $\Gamma$ of the knot equal to $\delta$.
The results for  $R_\mathrm{kpc}=1$ and $B_\mathrm{mG}=0.1$ are shown 
in Figure \ref{fig:beaming-spec} for $\delta = 3$.
As is seen, the overproduction of the optical flux can be 
avoided in this case as well.
The appropriate value of $K_p=10^{-2}$ \added{cm$^{-3}$} is similar to the non-beamed case,
so that the required proton power becomes as high as 
$10^{51} \,\, \text{erg} \,\, \text{s}^{-1}$, which seems to be unlikely.
However, we note that milder value of $\delta$ may reproduce the observed spectra , 
if the uncertainty of the ultraviolet flux exists by a factor of 1.5 or so.

\section{Conclusion}  \label{sec:conclusion}

We examined the  photo-hadronic  model of the X-ray emission from large scale jets 
of radio loud quasars specifically PKS 0637-752 and have found that Bethe-Heitler process is 
effective for the high energy electron/positron injection. 
Electrons and positrons from photo-pion production  mainly radiate
multi-TeV photons by synchrotron radiation and do not much contribute to X-ray emission.
For an appropriate choice of parameters such as $R=1$ kpc and $B=0.1$ mG, 
the required proton power is an order of $10^{49} \, \text{erg} \,\,  \text{s}^{-1}$,
\added{which is $\sim 120 L_\mathrm{Edd}$ for $M_\mathrm{BH} = 6.5 \times 10^8 M_\sun$},
when the energy density of protons is concentrated in the region 
of $\gamma_p =10^{9}$-$10^{10}$. 
Cooling tail of these electrons and positrons radiate optical synchrotron emission,
separate from the primary electrons. 
To avoid the overproduction of optical-ultraviolet flux,
either the energy distribution of the primary electrons has a super-exponential  cutoff or 
a mild degree of the relativistic beaming effect ($\delta \sim 3$) appears.
For the latter case, the required proton power tends to be large.
The Poynting and primary electron powers remain moderate. 
Because the value of the beaming factor is not strongly constrained
\citep[e.g.,][]{meyer2015}, further work is needed to determine
which mechanism is applicable to reduce the optical synchrotron emission.

Proton synchrotron radiation is a few orders of magnitude smaller than 
the photo-hadronic model prediction.
Thus, photo-proton model is an alternative option to 
explain the strong X-ray emission from large scale jets.

\added{The black hole mass of PKS0627-752 given by \cite{ljg2006} is 
$6.5 \times 10^8 M_\odot$.  
On the other hand,  \cite{gcj2001} gives $7.8 \times 10^9 M_\odot$.
The Eddington luminosity is 
$\sim 8.2 \times 10^{46}$ erg s$^{-1}$ and $\sim 9.8 \times 10^{47}$ erg s$^{-1}$ 
for $M_\mathrm{BH} = 6.5 \times 10^8 M_\odot$ and $7.8 \times 10^9 M_\odot$,
respectively.
Thus $10^{49}$ erg s$^{-1}$ is $\sim 10$ - $100$ times larger than the Eddington 
luminosity.
As for other AGNs, very luminous AGNs have been observed.
For example, \cite{gfv2009} showed that S5 0014+813 has
$\nu L_\nu \sim 10^{48}$ erg s$^{-1}$ in the optical and this corresponds to 
$0.17 L_\mathrm{Edd}$ for $M_\mathrm{BH} = 4 \times 10^{10} M_\sun$.
Some authors, e.g., \cite{sn2016}, on the other hand,  
performed radiation magnetohydrodynamical simulation of super-Eddington mass accretion.
In view of uncertainties of mass estimation and theoretical possibility 
of super-Eddington jet power, we believe our model is still viable, 
although the required proton power is very large.
}

Finally, our model predicts TeV emission by inverse Compton scattering of radio-optical
photons off pairs produced by the Bethe-Heitler process.
Emission by electrons/positrons produced by photo-pion processes will also 
contribute to TeV emission.

\acknowledgments
\added{We are grateful to the referee for useful comments that improved the manuscript 
considerably. }



\end{document}